\documentclass[12pt]{iopart}

\usepackage{iopams}
\usepackage{psfig}

\begin{document}

\jl{4}

\title[Geomagnetic effects on atmospheric \v{C}erenkov
images]{Geomagnetic effects on atmospheric \v{C}erenkov images}

\author{P M Chadwick, K Lyons, T J L McComb\footnote[1]{To whom
correspondence should be addressed.}, K J Orford,\\ J L Osborne, S M
Rayner, I D Roberts\footnote[2]{Now at Rolls Royce Plc., Derby, UK}, S E
Shaw\\ and K E Turver}

\address{Department of Physics, Rochester Building, Science
Laboratories, University of Durham, Durham DH1 3LE, UK}

\begin{abstract} 

Atmospheric \v{C}erenkov telescopes are used to detect electromagnetic
showers from primary gamma rays of energy $\sim 300$ GeV -- $\sim 10$
TeV and to discriminate these from cascades due to hadrons using the
\v{C}erenkov images. The geomagnetic field affects the development of
showers and is shown to diffuse and distort the images. When the
component of the field normal to the shower axis is sufficiently large
($> 0.4$ G) the performance of gamma ray telescopes may be affected,
although corrections should be possible. 

\end{abstract}


\submitted

\maketitle


\section{Introduction}

It is now 45 years since Cocconi \cite{kn:cocconi} drew attention to the
broadening effects of the geomagnetic field on the lateral development
of electron-photon cascades in the atmosphere. Allen \cite{kn:allen}
discussed the interaction of the geomagnetic field and the cascade
electrons/positrons as the origin of the radio frequency emission
produced by large cosmic ray showers. Earnshaw {\it et al}
\cite{kn:earnshaw} suggested that the geomagnetic separation of muons
could be used to estimate the height of origin of showers. Porter
\cite{kn:porter} and Browning and Turver \cite{kn:brover} made
calculations for the production of \v{C}erenkov radiation by
electron-photon cascades in the atmosphere, with allowance for the
effects of the geomagnetic field.

Ground based gamma ray astronomy exploits the magnifying effects of the
atmosphere to enable the detection of very low fluxes of gamma rays from
a variety of sources \cite{kn:fegan97}. The cascade of electrons
initiated by a primary gamma ray produces, at an altitude of about 10
km, \v{C}erenkov light which reaches the ground in a pool $\sim 300$ m
in diameter. It is possible to detect gamma rays of energy $>$ 300 GeV
via this \v{C}erenkov light produced in the atmosphere with an effective
collection area, for our Mark 6 telescope, of $5 \times10^{4}$ m$^{2}$.

Bowden {\it et al} \cite{kn:bowden91} discussed the effect of the
geomagnetic field on the performance of ground based gamma ray
telescopes. The interaction of the field and the cascade electrons
produces a broadening of the atmospheric \v{C}erenkov light image
resulting in a reduction in the density of light sampled by the
telescope; so the energy threshold for the telescope increases. This is
observed in the higher count rate for a telescope detecting showers
propagating along the lines of the field (with no spreading) than when
observing cascades developing perpendicular to the field lines (and
being spread), all other factors being the same. Typical differences in
measured count rate were about 20\% in these extreme cases. The
possibility was noted, on the basis of simulations, that an associated
rotation of the direction of the \v{C}erenkov light images may occur for
cascades developing under high magnetic fields and in unfavourable
directions. This could be of importance in ground-based gamma ray
studies since the orientation of the image in gamma ray initiated
cascades is the key to rejection of $>$ 99\% of the charged cosmic ray
background \cite{kn:hillas85}.
 
Lang {\it et al} \cite{kn:lang93} showed that the measurements of TeV
gamma rays from the Crab nebula using the imaging \v{C}erenkov technique
were not significantly affected when the magnetic field was $<$ 0.35 G.

We report measurements made using a ground based gamma ray telescope of
cascades developing under the influence of fields up to 0.55 G.
Observations with the Mark 6 telescope operating in Narrabri, Australia
which is discussed by Armstrong {\it et al} \cite{kn:armstrong} are
subject to such magnetic fields when observing objects to the south. The
observational data demonstrate all the effects of the geomagnetic field
on cascades which are predicted by simulations.


\section{Geomagnetic effects on EM showers}

\subsection{Simple model}

The lateral distribution of electrons and positrons near the maximum of
an electromagnetic shower in air is approximately axially symmetric. The
presence of a transverse component of magnetic field will cause the
electrons and positrons to separate and the lateral distribution to
become wider along the radial direction perpendicular to the transverse
field component. The order of magnitude of this effect may be estimated.

A relativistic electron of energy $E$ (MeV) will follow a circular path
in a transverse magnetic field of strength $H$ (G) with a radius of
curvature $R$ (km) of $\sim{E}/(30 H)$. The energy of the typical
\v{C}erenkov light-producing electron at shower maximum is $\sim$ 100
MeV (the critical energy in air), which is at a height of $\sim$ 10 km
for a 250 GeV primary gamma ray. The radius of curvature of these
electrons for a maximum transverse magnetic field of 0.56 G is $\sim$ 6
km. The typical length of such an electron's path in air at 10 km
altitude is $\sim$ 1 km, giving a lateral deflection of $\sim$ 80 m. At
a height of 10 km this corresponds to an angular deflection of $\sim 0.5
^{\circ}$, the same order as the Coulomb scattering width. It should be
possible to detect this as a broadening of the shower along one axis.

The effect of the broadening will depend on the orientation of the image
with respect to the direction of the magnetic field. The direction of
the shower axis in the atmosphere is within $\sim$1$^{\circ}$ of the
telescope-source line. The transverse magnetic field component therefore
lies very close to the plane of the focal image of the shower. The
parameters of an image which are used to discriminate between gamma rays
and hadrons depend primarily on the second spatial moments. In a
coordinate frame in which the x-axis is aligned with the projected field
direction, these are $\sigma _{\rm x}^{2}$ , $\,\sigma _{\rm y}^{2}$ and
$\sigma _{\rm xy}$. The orientation angle of the long axis of the image
is \mbox{$\tan ^{-1} \left[\left(\sigma _{\rm y}^{2}-\sigma _{\rm x}^{2}
+ \sqrt{(\sigma _{\rm y}^{2}-\sigma _{\rm x}^{2})^{2}+4\sigma _{\rm
xy}^{2}}\right)/(2\sigma_{\rm xy})\right]$}. Those shower images whose
major axes are aligned with the projected direction of the magnetic
field have \mbox{$\sigma _{\rm xy}=0$}. The moment $\sigma _{\rm y}^{2}$
will be expected to be increased by the magnetic field and the images
will be widened only. For images at other orientations, \mbox{$\sigma
_{\rm xy} \neq 0$} and the result of an increase in $|\sigma _{\rm xy}|
$ will be to rotate their minor axes towards the projected field
direction. The main parameter used to discriminate gamma rays is {\it
ALPHA}, the angle contained within the long axis of the image and the
radius vector from the source position to the centroid of the image. Any
rotation of the image may affect {\it ALPHA} and hence the sensitivity
for gamma ray detection. This is discussed further in
section~\ref{simulations}.


\subsection{Simulations}\label{simulations}

Monte Carlo simulations of gamma ray and hadron cascades have been made
to investigate the strength of this geomagnetic effect. These
simulations have been performed using a Monte Carlo code developed from
that used in our earlier work \cite{kn:bowden91} which incorporates the
responses of the University of Durham telescopes. The results of the
simulations have been validated against established simulation codes
e.g. MOCCA \cite{kn:hillas85}. Agreement has been found in all cases.

\begin{figure}[tb]

\centerline{\psfig{file=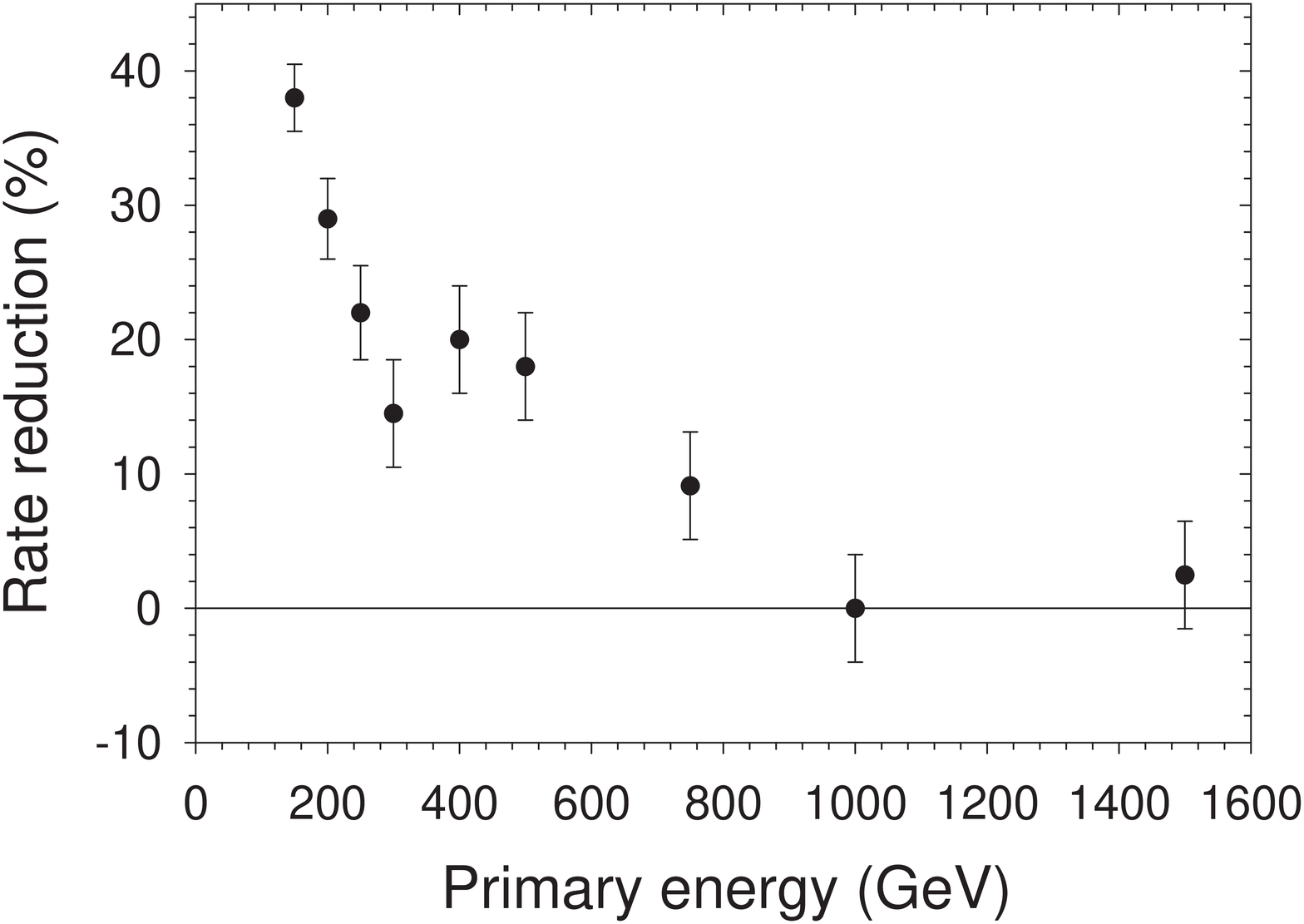,height=6cm}}

\caption{The simulated effect of the geomagnetic field on the brightness
of the observed image. We show the \% reduction in event rate for
hadrons as a function of primary energy for a field of 0.5 G relative to
the zero-field case.} \label{fig:simtotal}

\end{figure}

The consequence of an increase in the lateral and angular spread of the
shower electrons is a decrease in the density of \v{C}erenkov light on
the ground. This leads to a decrease in counting rate for fixed
threshold, which was reported by us from measurements, using a
non-imaging telescope, of counting rate against azimuth at fixed zenith
angle \cite{kn:bowden91}. Figure~\ref{fig:simtotal} shows the simulated
reduction in the counting rate of a gamma ray telescope of the type used
by us for hadron primaries, relative to the zero-field case, caused by a
transverse magnetic field of 0.5 G. This is consistent with our earlier
measurements --- a reduction of up to $\sim 20\%$ in count rate for
gamma ray energies $\geq 300$ GeV \cite{kn:bowden91} and, apparently, a
greater reduction for $< 300$ GeV cascades.

\begin{figure}[tb]

\centerline{\psfig{file=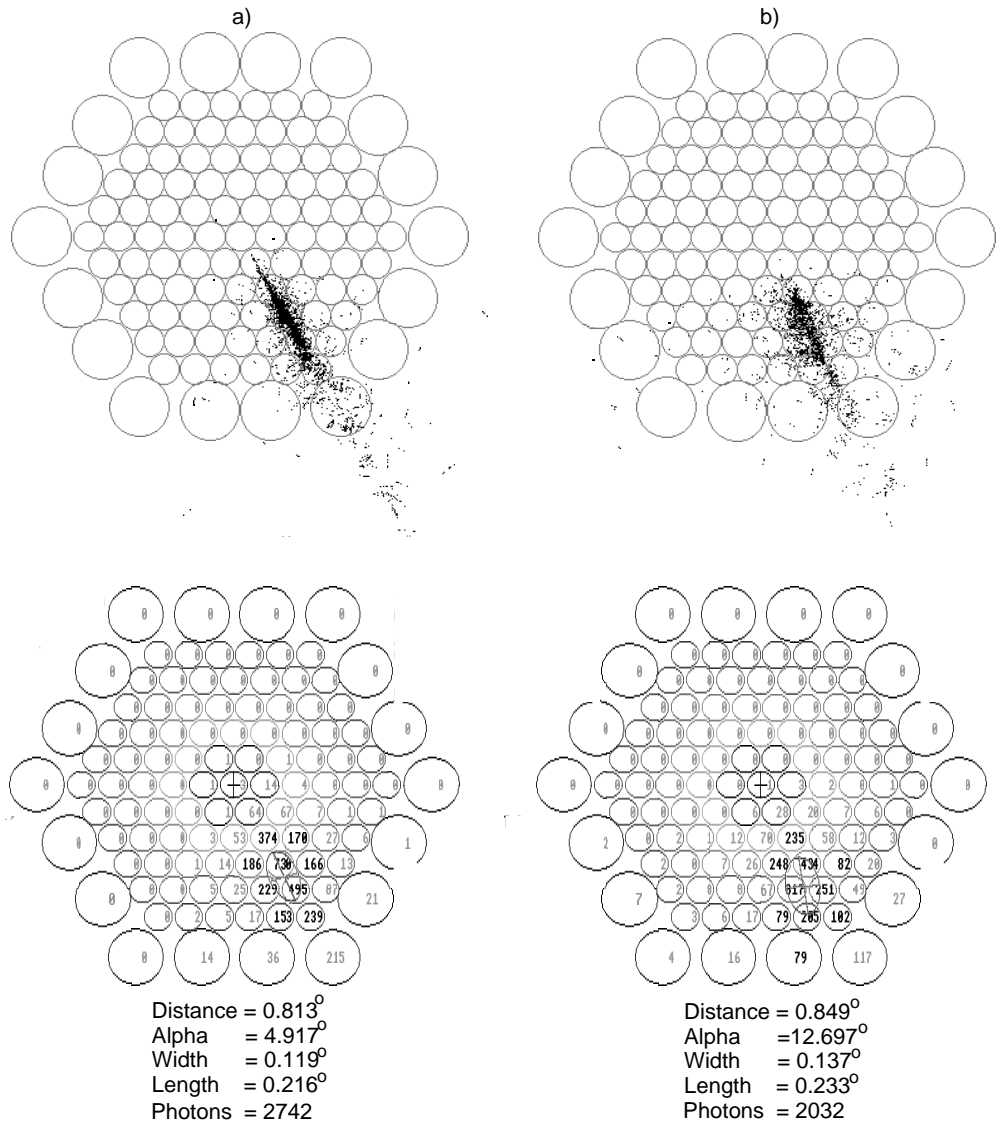,height=15cm}}

\caption{Simulations of a 1 TeV gamma ray cascade. The simulation has
been performed with geomagnetic field values of (a) 0 and (b) 0.56 G.
For each image, both a raw photon map (upper plot) and the results of
fitting standard image parameters after allowance for mirror defects,
pixelation and noise (lower plot) are shown.} \label{fig:simimage}

\end{figure}

In our earlier work \cite{kn:bowden91} we suggested that the
distribution of recorded photons in an imaging camera may be distorted.
We show in figure \ref{fig:simimage} the pattern of photons in
individual simulated 1 TeV gamma ray cascades developing under the
effects of 0 and 0.56 G geomagnetic fields. We also show the predicted
responses of our camera after allowing for optical distortion,
pixellation of the camera, measurement noise and passing through our
standard data analysis package. The simulations for the pairs of cascades
developing under the influence of a field and no field have been
initiated using the same random number seed; this ensures, as far as is
possible, that the early development of the cascade (which dominates the
resulting image and is nearly independent of field) is identical for the
two simulations. The expected effects on the shape of the image
(widening or lengthening and rotation, depending on the orientation of
the image with respect to the magnetic field direction) was apparent,
both in the distribution of photons hitting the detector and in the
processed image.

\begin{figure}[bt]

\centerline{\psfig{file=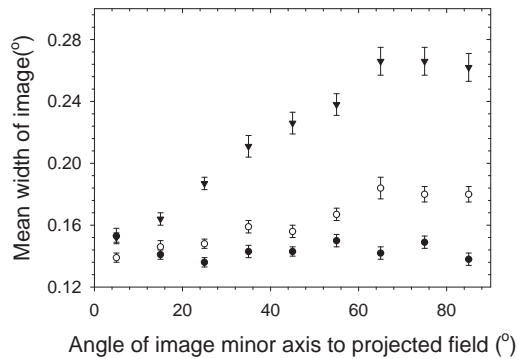,height=6cm}}

\caption{The effect of the geomagnetic field on the width of simulated
500 GeV gamma ray shower images which are not subjected to any
broadening effects of the mirror or camera. $\fullcircle$ is for a
geomagnetic field of 0.0 G, $\opencircle$ for 0.2 G and
$\blacktriangledown$ for 0.5 G.} \label{fig:simswid}

\end{figure}

The predicted width of a shower image was investigated as a function of
magnetic field strength and angle for gamma rays of energy 500 GeV.
Results are shown in figure~\ref{fig:simswid} for transverse field
strengths of 0.0, 0.2 and 0.5 G. No allowance is made for the broadening
effects of the defects in the mirror or the pixellation of the recording
PMT camera. The effect of a 0.5 G field on the width (minor axis) of
such undistorted gamma ray images is seen to be greatest when the image
minor axis is perpendicular to the projected field direction, as
expected.

\begin{figure}[tb]

\centerline{\psfig{file=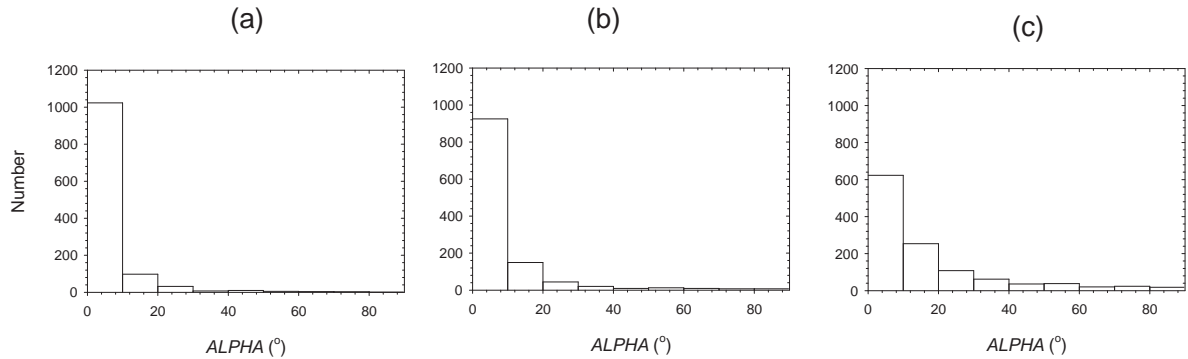,height=8cm}}

\caption{The effect of the geomagnetic field on the distribution in
pointing angle {\it ALPHA} for simulated 500 GeV gamma ray shower
images. (a) is for a geomagnetic field of 0.0 G, (b) for 0.2 G and (c)
for 0.5 G. Again, there is no allowance for the effects of mirror optics
or camera pixellation.} \label{fig:simsalpha}

\end{figure}

Small transverse fields of 0.2 G have a limited effect on the image
width. However, the effect on the pointing angle {\it ALPHA} become
noticable and may vary both with the image's angle to the magnetic field
and with its eccentricity (defined as the ratio {\em WIDTH}/{\em
LENGTH}). For example, an image with very low eccentricity may be
widened only by a small amount but rotated through a large angle.
Histograms in {\it ALPHA} are shown in figure~\ref{fig:simsalpha} for
the simulated gamma rays, for which widths were shown in
figure~\ref{fig:simswid}, under the influence of fields of 0.0, 0.2 and
0.5 G.

The effects of mirror quality and camera pixellation of a typical
telescope may reduce this change in width but may not mask the change in
orientation of the image.


\subsection{Magnetic fields appropriate to Narrabri observations}

\begin{figure}[tb]

\centerline{\psfig{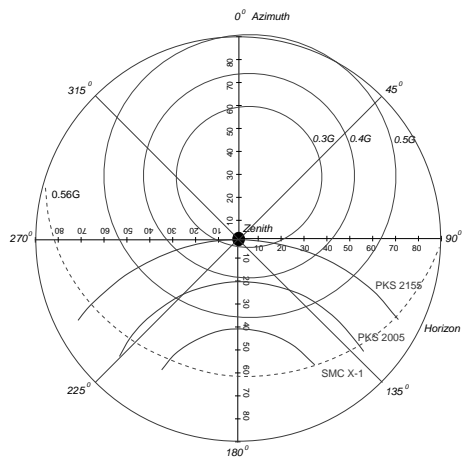}}

\caption{The tracks for various potential gamma ray sources plotted to
show the geomagnetic fields experienced at Narrabri, Australia.}
\label{fig:tracks}

\end{figure}

We show in figure~\ref{fig:tracks} the celestial sphere, as viewed from
the site of the Mark 6 telescope in Narrabri, Australia, showing the
loci of constant magnetic field strength. We indicate the tracks on the
sky for the telescope during observations of a number of potential gamma
ray sources which have been intensively studied. In fact, most of the
sources studied at Narrabri are in the azimuth range $135^\circ -
225^\circ$ and in directions for which the magnetic field is $>$ 0.35 G
and in many cases $>$ 0.5 G. The telescope sensitivity is therefore
likely to be reduced by the broadening of the distribution of {\it
ALPHA} for many of our observations, unless corrections are applied.


\section{The equipment}

All data reported here have been obtained with the Mark 6 VHE gamma ray
telescope which has been described in detail by Armstrong {\it et al}
\cite{kn:armstrong}. The telescope measures gamma rays in the energy
range 300 GeV -- 10 TeV, with a 50\% trigger probability at an energy of
$\sim 300$ GeV. It comprises an alt-azimuth mount with three 7m diameter
f/1.0 aluminium mirrors with an rms spread for a point source of
$0.18^\circ$. The pixel size for a $1"$ diameter photomultiplier is
$0.25^\circ$. The focal plane of the central mirror contains a 109-pixel
camera comprising a close-packed hexagonal array of 91 1-inch diameter
photomultipliers surrounded by a ring of 18 2-inch diameter
photomultipliers. The field of view of the imaging camera is $3^\circ$
wide. The mirrors on the left and right of the mount each contain a
triggering detector of 19 2-inch photomultipliers in a close-packed
hexagonal array. Each photomultiplier is contained in a mumetal shell
designed to give magnetic and electrostatic shielding. The effect of the
geomagnetic field on the performance of the photomultipliers, with and
without shielding, has been studied \cite{kn:roberts}. For the 1-inch
Hamamatsu R1924 PMT the effect of a 0.5 G field on the gain was not
measureable and less than 1\%. For the 2-inch Phillips XP3422 PMT the
measured decrease in gain due to a 0.5 G field was 1\%. Shielded
photomultipliers of both types show no measurable effects due to the
magnetic field.

The control of the attitude of the telescope is via DC servomotors
driving onto gears mounted directly on the telescope structure. Angles
are sensed by absolute digital 14-bit shaft encoders with a resolution
of $0.022^{\circ}$. The calculated azimuth and zenith of a source is
compared by a digital servomechanism (employing 12-bit resolution) with
the shaft encoder outputs at 100 ms intervals. The error signals are
passed via DACs to the DC motor amplifiers. These provide damping on
acceleration and stabilise the movements of the telescope structure.
Thus although the telescope pointing is known to a resolution of
$0.022^\circ$, the source can be offset from the camera centre by up to
$0.1^\circ$ by this mechanism.

The attitude of the telescope is measured in two ways. The shaft encoder
positions are recorded for each event to 14-bit accuracy. This gives a
measurement of the pointing to $\pm 0.022^{\circ}$ within the telescope
system. In addition, a coaxial optical CCD camera is mounted on the
telescope. The output of this CCD camera is continuously monitored by
microcomputer, which measures the position and brightness of a nominated
guide star within the $ 2^\circ \times 2^\circ$ field. This information
is integrated into the data stream on an event-by-event basis. Guide
stars of magnitude $m_{\rm{v}} \leq 6$ can be employed, providing
absolute position sensing for the telescope to better than
$0.008^{\circ}$.

False source analysis has been shown to be a useful method of
demonstrating that $\gamma$-ray like events originate from the source
direction \cite{kn:kifune1995}. Conversely, such analyses may also be
used to verify the steering performance of a telescope, if an
established $\gamma$-ray emitter is observed. However, we have only one
data set for a gamma ray source containing data which are {\em not}
subject to the effects of a strong geomagnetic field --- see section
\ref{pks2155alpha}.


\section{Experimental results}

\subsection{Data}

Most of the data considered here were taken during routine observations
of potential gamma ray sources during 1996 -- 1998. A small amount of
data was taken in dedicated measurements at fixed values of azimuth and
zenith, corresponding to a range of values of the geomagnetic field. The
only restriction imposed on data was that events should lie within
$1^\circ$ of the centre of the camera (to avoid edge effects) and were
large enough (5 times the triggering threshold) to ensure that their
shape was well measured.

\subsection{ Reduction in image brightness}

The investigation of the sensitivity of the count rate to the magnetic
field which established the effect with the Mark 3 and Mark 4 telescopes
\cite{kn:bowden91} has been repeated for the Mark 6 telescope. We find
that the count rate observed with the Mark 6 telescope for cascades
developing under minimum and maximum values of the geomagnetic field at
a constant zenith angle of $40^\circ$ differs by about 15\%, as
expected.

\subsection{Effects on the shape of \v{C}erenkov images of hadron
showers}

\begin{figure}[tb]

\centerline{\psfig{file=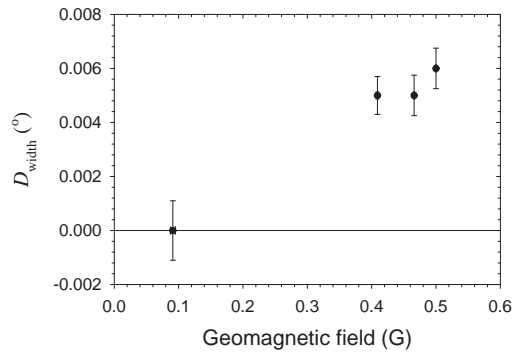,height=6cm}}


\caption{The observed difference in the mean width ($D_{\rm width}$) for
images in hadronic cacades developing parallel and perpendicular to
fields of varying strength for zenith angles in the range $35^\circ -
40^\circ$. The concentration of data at values of high magnetic field
strength reflects the directions of observation of potential VHE sources
from Narrabri.} \label{fig:distortion}

\end{figure}

The widths of images recorded by the Mark 6 telescope in directions
parallel and perpendicular to the magnetic field have been investigated.
For a large sample of cascades measured during observations of a range
of sources the differences of the mean width in the parallel and
perpendicular directions are shown in figure \ref{fig:distortion} for a
range of values of distorting field. The widths of images in directions
perpendicular to the field are significantly greater than those in
parallel directions for fields $> 0.4$ G. Although the differences in
widths are small, because of the effects of pixellation and noise which
are common to all data, the measurements are free from systematic
effects. This is because the orientation of images to the magnetic field
depends on the orientation of the image in the camera. Measurements of
the width of the images parallel and perpendicular to the magnetic field
are derived from events distributed throughout the same observation. 

We have attempted to measure the differences in widths of images in
cascades measured at fixed zenith angles of $40^\circ$ for values of
fields less than and greater than 0.35 G. Such measurements are difficult
and require large exposures ($\gtrsim 100$ hr) to give significant
differences. For example, with exposures $\sim 1$ hr we find for $B <
0.35$ G, the mean width is $0.292^\circ \pm 0.0035^\circ$ (SD $=
0.041^\circ$) and for $B > 0.35$ G the mean width is $0.283^\circ \pm
0.0036^\circ$ (SD $= 0.042^\circ$). The small difference in means,
significant at the $\sim 2 \sigma$ level, demonstrates the difficulty in
making such a small measurement for which the standard deviation is
approximately similar for all $B$.

\subsection{Effects on the orientation of \v{C}erenkov images of hadron
showers}

In the absence of any magnetic field and detector triggering biases, the
distribution of the orientation of the images in the camera of a
telescope will be isotropic. If we define a coordinate system where the
angle $a_{\rm x}$ is the angle of the long axis of an image to the
horizontal in the camera frame, then the distribution of $a_{\rm x}$
should be flat between $-90^{\circ}$ and $90 ^{\circ}$. Minor deviations
from uniformity in angle might be expected near to threshold because of
the increased effect of small changes in triggering probability and the
six-fold symmetry of the hexagonal arrangement of close-packed
detectors.

\subsubsection{Distribution of projected shower angles}

\begin{figure}[tb]

\centerline{\psfig{file=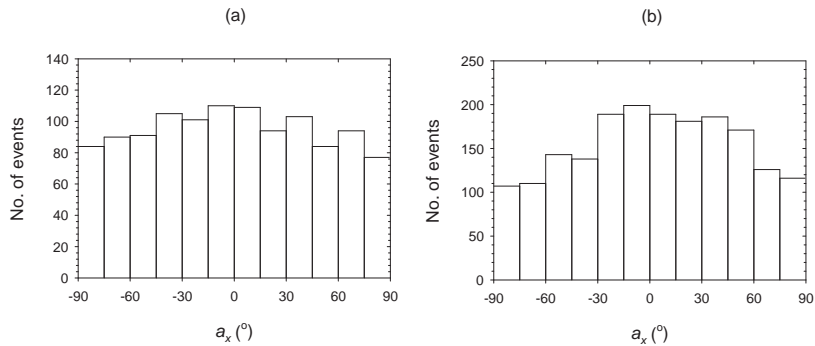,height=6cm}}

\caption{Distribution of $a_{\rm x}$ for cascades developing under (a) a
small transverse geomagnetic field ($B = 0.15$ G) and (b) a
large transverse geomagnetic field ($B = 0.52$ G). \label{fig:aysmallh}}

\end{figure}

We show in figure~\ref{fig:aysmallh}(a) the distribution in the angle
$a_{\rm x}$ for background hadronic events which are subject to small
values of the transverse magnetic field (0.15 G). We show in
figure~\ref{fig:aysmallh}(b) the distribution in $a_{\rm x}$ for images
due to hadron-induced cascades recorded in special observations at fixed
azimuth and zenith angles which gave a transverse geomagnetic field of
0.52 G. All of these events were recorded in dedicated observations with
the telescope at a fixed zenith angle ($40^\circ$). Data at different
values of the transverse magnetic field were obtained by varying the
azimuth angle. All data presented here were recorded within a period of
one hour, so potential variations due to changes in camera performance,
atmospheric clarity, etc. were minimised. The data were processed
following our normal procedures (see e.g. \cite{kn:chadwick98b}). The
requirement that images have a minimum brightness and fall within the
camera ensures that the data are free of effects of variations in night
sky brightness. The data for $B = 0.15$ G show the expected distribution
indicative of a near isotropic distribution of directions in the camera
with a small peak. We note a strong anisotropy, with a peak containing
twice the number of events resulting from the skewing or distortion of
the image, for events subject to a $0.5$ G field. The maximum of the
distortion occurs at $a_{\rm x} = 0$, which is appropriate for those
observations which were made at magnetic south.

\subsubsection{Correlation of amplitude of $a_{\rm x}$ peak and geomagnetic
field strength}

\begin{figure}[tb]

\centerline{\psfig{file=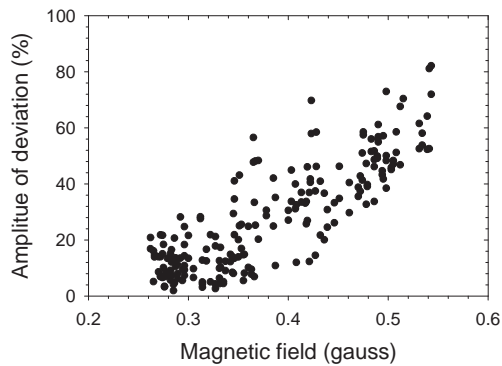,height=6cm}}

\caption{The measured amplitude of $a_{\rm x}$ anisotropy as a function of
transverse geomagnetic field.} \label{fig:ayampvh}

\end{figure}

The magnitude of the anisotropy displayed in
figure~\ref{fig:aysmallh}(b) should depend on the strength of the
magnetic field. In figure~\ref{fig:ayampvh} we plot the amplitude of the
peak of the distortion in the $a_{\rm x}$ distribution as a function of
the magnetic field strength. Each point corresponds to the result for a
15 minute segment of data.

%
%

Note that for values of transverse component of the geomagnetic field
less than 0.35 G there is no great distortion of the $a_{\rm x}$
distribution, as suggested by the work of Lang {\it et al}
\cite{kn:lang93}, but for values of field in excess of 0.4 G substantial
distortion occurs.

\subsubsection{Correlation of directions of $a_{\rm x}$ peak and
geomagnetic field}

\begin{figure}[tb]

\centerline{\psfig{file=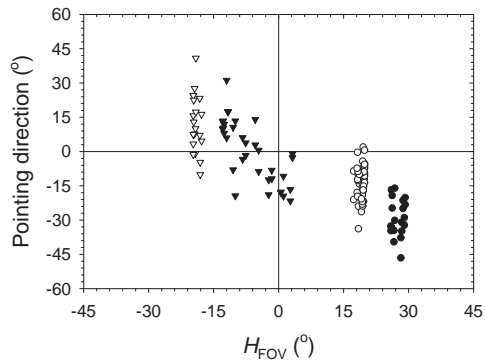,height=6cm}}

\caption{The measured direction of $a_{\rm x}$ anisotropy as a function
of the direction of transverse geomagnetic field. $\opentriangledown$
represents data from the directions of Cen X-3, $\blacktriangledown$
from SMC X-1, $\opencircle$ from PKS 2005--489 and $\fullcircle$ from
PKS 2155--304 to cover a range of $H_{\rm FOV}$.} \label{fig:aydirvh}

\end{figure}

The position of the peak in the $a_{\rm x}$ distribution depends on the
angle between the projected magnetic field and the vertical direction in
the camera --- $H_{\rm FOV}$. We show in figure~\ref{fig:aydirvh} the
correlation between $H_{\rm FOV}$ and the position of the peak in
$a_{\rm x}$. The data demonstrate the expected relation between these
angles. Again, each point is based on a 15 min sample of data taken
during routine observations of a range of potential gamma ray sources.

\subsection{Summary}

We have evidence that the geomagnetic field should, and does, influence
the lateral development of atmospheric cascades. The observed effects on
the threshold of a telescope and on the shape of the image and the
magnitude and phase of the distortion of the pointing of images produced
by background cosmic ray protons are as expected.


\section{The orientation of gamma ray images}

A number of potential gamma ray sources has been observed using the Mark
6 telescope at Narrabri. In several cases there is evidence for gamma
ray emission \cite{kn:chadwick98a,kn:chadwick98b,kn:chadwick99}. The
evidence comes mainly from a comparison of the {\it ALPHA} distributions
for data selected on the basis of image shape for the ON-source and
OFF-source scans. The difference between the {\it ALPHA} distributions
should show an excess of events --- the gamma ray candidates --- at low
values of {\it ALPHA}. If the data were taken in geomagnetically
unfavourable directions, as is the case for most of our data, it might
be expected that the {\it ALPHA} distribution of the excess events would
be wider than that for data taken in more favourable geomagnetic
directions. Consideration of data recorded from a gamma ray source which
are subject to $B < 0.35$ G may provide the only true indication of the
{\it ALPHA}-distribution for our telescope.

\subsection{Observations of PKS 2155--304}\label{pks2155alpha}

\begin{figure}[tb]

\centerline{\psfig{file=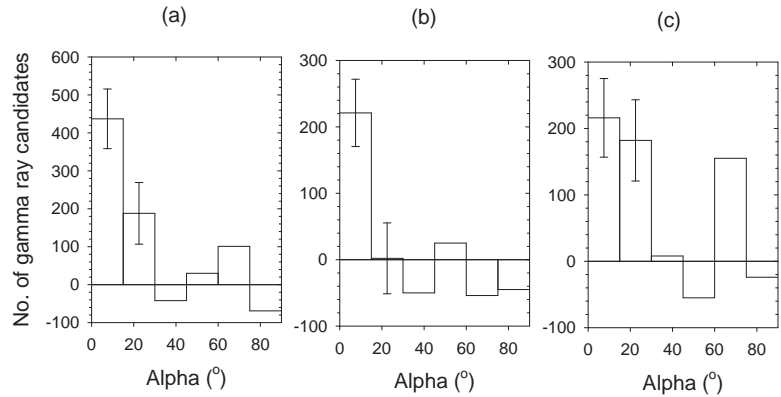,height=6cm}}

\caption{The distribution in {\it ALPHA} of excess gamma ray events from
PKS 2155--304. (a) is for all events, (b) those for transverse fields
$<$ 0.35 G. and (c) those for transverse fields $>$ 0.35 G.}
\label{fig:alphaplot2}

\end{figure}

Observations of this object were made with the cascades recorded over a
range of transverse geomagnetic field strengths between 0.25 and 0.5 G.
The total data set contained 41 hrs of observation, as reported
\cite{kn:chadwick99}. The {\it ALPHA} plot for the difference between
the ON-source and OFF-source data taken at zenith angles $\theta <
45^{\circ}$ is shown in figure~\ref{fig:alphaplot2}(a). The significance
of the excess at {\it ALPHA} $< 30^\circ$ is $5.7 \sigma$ according to
the maximum likelihood method \cite{kn:gibson82}.

We are able to select a subset of data for which the strength of the
projected field to which the cascades were exposed was $<$ 0.35 G and
for which minimal distorting effect would be expected. The {\it ALPHA}
plot for this subset is shown in figure~\ref{fig:alphaplot2}(b). The
excess events all have {\it ALPHA} $ < 15^\circ$ (significance $4.7
\sigma$). The distribution is narrower than that of the total dataset
and is typical of that expected for a $0.25 ^{\circ}$ pixel camera. (In
the absence of any other data for gamma rays detected with our telescope
and {\it not} subject to the effects of the magnetic field, we assume
that this is reasonable.)

The {\it ALPHA} plot for the majority of the events for which the field
is $>$ 0.35 G, is shown in figure~\ref{fig:alphaplot2}(c). It is evident
that the width of the peak is larger for these events recorded under the
influence of higher transverse magnetic fields, with equal populations
for values of {\it ALPHA} between $0^\circ$ -- $15^\circ$ ($3.6 \sigma$)
and $15^\circ$ -- $30^\circ$ ($3.1 \sigma$). It should be noted that the
peak at $60^\circ$ -- $70^\circ$ is superimposed on an {\em ALPHA}-plot
with increasing frequency for {\em ALPHA} approaching $90^\circ$ --- see
\cite{kn:chadwick99}. The significance of this peak is therefore $\sim 2
\sigma$ (before allowing for the number of bins in the {\em
ALPHA}-plots).

On the basis of our only measurement of a gamma ray source not subject
to the effects of the geomagnetic field, we have evidence that the
excess events with $15^\circ <$ {\em ALPHA} $< 30^\circ$ are confined to
those measurements made with fields $> 0.35$ G. This conclusion is
significant at the 2.5\% level, on the basis of a $2 \times 2$
contingency test of the populations of the $15^\circ - 30^\circ$
bins in figures~\ref{fig:alphaplot2}(b) and \ref{fig:alphaplot2}(c).

\section{Conclusions}

We have demonstrated that, on the basis of simulations and measurement,
the \v{C}erenkov images from gamma rays and cosmic rays are
broadened and rotated by the geomagnetic field. The broadening results
in an increase in the telescope threshold and a reduction in counting
rate. The rotation of the images away from the projected direction of
the magnetic field in the image plane broadens the {\it ALPHA}
distribution.

These results suggest that the geomagnetic field can have an important
effect on the operation of atmospheric \v{C}erenkov telescopes in some
directions and that, for detected and candidate sources, our Narrabri
site is particularly susceptible. These effects may be removed using
appropriate correction techniques; unlike noise, geomagnetic effects do
not reduce the information contained in the image.

\ack

We are grateful to the UK Particle Physics and Astronomy Research
Council for support of the project and the University of Sydney for the
lease of the Narrabri site. The Mark 6 telescope was designed and
constructed with the assistance of the staff of the Physics Department,
University of Durham.


\section*{References}

\begin{thebibliography}{99}

\bibitem{kn:cocconi}Cocconi G 1954 \PR {\bf 93} 646

\bibitem{kn:allen}Allan H R 1970 {\it Acta. Phys. Hung. Suppl.} {\bf 29
(3)} 699

\bibitem{kn:earnshaw}Earnshaw J R, Machin A C, Orford K J, Pickersgill D
R and Turver K E 1971 {\it Proc. 12th Int. Conf. Cosmic Rays, Hobart}
{\bf 3} 1081

\bibitem{kn:porter}Porter N A 1973 {\it Nuovo Cimento Lett.} {\bf 8} 481

\bibitem{kn:brover}Browning R and Turver K E 1977 {\it Nuovo Cimento A}
{\bf 38} 223

\bibitem{kn:fegan97}Fegan D J 1997 \jpg {\bf 23} 1013

\bibitem{kn:bowden91}Bowden C C G, Bradbury S M, Chadwick P M, Dickinson
J E, Dipper N A, Edwards P J, Lincoln E W, McComb T J L, Orford K J,
Rayner S M and Turver K E 1992 \jpg {\bf 18} L55

\bibitem{kn:hillas85}Hillas A M 1985 {\it Proc. 19th Int. Cosmic Ray
Conf.} {\bf 3} 445

\bibitem{kn:lang93}Lang M J, Akerloff C W, Cawley M F, Chantell M, Fegan
D J, Gillanders G H, Hillas A M, Lamb R C, Lewis D A, Meyer D I and
Weekes T C 1994 \jpg {\bf 20} 1841

\bibitem{kn:armstrong}Armstrong P \etal 1999 {\it Experimental Astron.}
in press

\bibitem{kn:roberts}Roberts I R 1999 {\it PhD Thesis} (University of
Durham)

\bibitem{kn:kifune1995}Kifune T {\it et al} 1995 {\it Astrophys. J.}
{\bf438} L91

\bibitem{kn:chadwick98a}Chadwick P M, Dickinson M R, Dipper N A, Holder
J, Kendall T R, McComb T J L, Orford K J, Osborne J L, Rayner S M,
Roberts I D, Shaw S E and Turver K E 1998 {\it Astroparticle
Phys.} {\bf 9} 131

\bibitem{kn:chadwick98b}Chadwick P M, Dickinson M R, Dipper N A, Kendall
T R, McComb T J L, Orford K J, Osborne J L, Rayner S M, Roberts I D,
Shaw S E and Turver K E 1998 {\it Astrophys. J.} {\bf 503} 391

\bibitem{kn:chadwick99}Chadwick P M, Lyons K, McComb T J L, Orford K J,
Osborne J L, Rayner S M, Shaw S E, Turver K E and Wieczorek G J 1999
{\it Astrophys. J.} {\bf 513} 161

\bibitem{kn:gibson82} Gibson A I, Harrison A B, Kirkman I W, Lotts A P,
Macrae J H, Orford K J, Turver K E and Walmsley M 1982, {\it Proc. Int.
Workshop on Very High Energy Gamma Ray Astronomy}, ed P V Ramana Murthy
and T C Weekes, (Bombay: Tata Institute) p 97

\end{thebibliography}
\end{document}